\documentclass{aastex}

\usepackage{graphicx}
\usepackage{amssymb}
\usepackage{amsmath}
\usepackage{xcolor}
\usepackage{bm}
\usepackage{xspace}
\usepackage{pgffor}
\usepackage{upgreek}
\usepackage{nicefrac}
\usepackage[normalem]{ulem}
\usepackage[single=true]{acro}

\newcommand{\be}{\begin{equation}} 
\newcommand{\ee}{\end{equation}} 
\newcommand{\nn}{\mbox{} \nonumber \\ \mbox{} }
\newcommand{\ba}{\begin{eqnarray}}
\newcommand{\ea}{\end{eqnarray}}
\newcommand{\om}{\omega}
\newcommand{\Alfven}{Alfv\'{e}n }

\newcommand\eg{\textit{e.g.,\ }}

\newcommand{\Bf}{{magnetic field}}
\newcommand{\Bfs}{{magnetic fields}}
\newcommand{\Ef}{{electric  field}}
\newcommand{\Efs}{{electric fields}}

\newcommand{\NS}{neutron star}

\newcommand{\WD}{{white dwarf}}

\newcommand{\ms}{magnetosphere}
\newcommand{\mss}{magnetospheres}

\newcommand{\Lf}{Lorentz factor}
\newcommand{\Lfs}{Lorentz factors}

\begin{document}

\title{Magnetospheric interaction in white dwarf binaries  AR Sco  and  AE Aqr}
\author
%[Lyutikov \etal]
{
 Maxim Lyutikov $^1$,
 % \thanks{Correspondence author: lyutikov@purdue.edu (ML)}
 Maxim Barkov $^{1,2}$, 
 Matthew Route $^3$, 
 Dinshaw Balsara $^4$,
 Peter Garnavich  $^4$ and Colin Littlefield $^4$\\
  $^1$ Department of Physics and Astronomy, Purdue University, West Lafayette, IN 47907-2036, USA\\
  $^2$ Astrophysical Big Bang Laboratory, RIKEN, 2-1 Hirosawa, Wako, Saitama 351-0198, Japan\\
  $^3$ Research Computing, Purdue University, 155 S. Grant St., West Lafayette, IN 47907-2036, USA\\
  $^4$ Department of Physics, University of Notre Dame, Notre Dame, IN 46556, USA
}

%\author[Lyutikov]{
% Maxim Lyutikov, Maxim Barkov, Daniel Proga\\
 % \thanks{Correspondence author: lyutikov@purdue.edu},\\
% Department of Physics and Astronomy, Purdue University, West Lafayette, IN 47907-2036, USA\\
%}

%\author{Maxim Lyutikov\\ Department of Physics and Astronomy, Purdue University,  525 Northwestern Avenue, West Lafayette, IN 47907-2036 }

%\item MR: is there a  mechanism to link ORCIDs to the author list? 

%\maketitle

\begin{abstract}
We develop a model of   the white dwarf (WD) - red dwarf  (RD) binaries    AR Sco  and AE Aqr as systems in a 
transient propeller stage of highly asynchronous intermediate polars. The WDs are  relatively weakly magnetized  with \Bf\ of   $\sim 10^6$ G. We explain  the salient observed  features of the systems  due to the 
magnetospheric interaction of two stars. Currently, the WD's spin-down is determined by the mass loading of the WD's \ms\ from the RD's at a mild rate of  $\dot{M} _{WD} \sim  10^{-11} M_\odot $/yr. Typical loading distance is determined by   the ionization of the RD's wind by the WD's UV flux.
 The WD was previously spun up by a period of high accretion rate from the RD via Roche lobe overflow with    $\dot{M} \sim  10^{-9} M_\odot $/yr, acting for as short a period  as tens of thousands of years.  The   non-thermal  X-ray and  optical  synchrotron emitting  particles originate  in reconnection events in  the  \ms\ of the WD due to the interaction with  the  flow from the RD.
  In the case of AR Sco, the reconnection events produce signals at the WD's  rotation and beat periods -  this modulation is due to the 
  changing relative  orientation of the companions' magnetic moments and resulting variable reconnection conditions.   Radio emission  is produced in the \ms\ of the RD,   we hypothesize, in a way that it is physically similar to the Io-induced   Jovian decametric  radiation. 
\end{abstract}

\maketitle

\section{Introduction}
%\section{: Magnetized, Rapidly Rotating White Dwarfs in Close Binaries }
%\subsection{Cataclysmic variables}

Cataclysmic variable stars (CVs) are interacting binary systems where a low-mass donor star transfers mass to a white dwarf (WD) \citep{2003cvs..book.....W}. 
CVs can lead to a variety of astrophysical phenomena   that range from powerful thermonuclear explosions, to the generation of non-thermal radio and high energy emission, and emission  of low frequency gravitational waves  that may be detectable by the \textit{LISA} mission \citep{2019BAAS...51c.168T}.
 Such systems are  formed when the more massive component in a stellar binary expands towards the end of its stellar life and engulfs its companion;  this brief and dynamically violent common envelope phase shrinks the orbital separation, and results in a radically different evolution compared to single star evolution. 
In the resulting compact binary, gas flows from the donor star to the WD.
The accretion of this gas onto the WD results in variability over a range of timescales, from seconds to months.

A magnetized WD (mWD) adds  another dimension to the mass exchange as the field can directly channel material to the vicinity of the mWD magnetic poles, speeding the release of gravitational energy and generating strong non-thermal emission. 
%Unless the mWD has a sufficiently high field strength, its spin period will be different than the binary orbital period. The interaction between the spin and orbit can generate a beat period and its harmonics. 
CVs can then be divided into magnetic and non-magnetic CVs, with the former further divided into polars and intermediate polars.  These systems were identified by the linear or circular polarization of their optical light that varied with the binary orbital period, as found in the prototypical system, AM Her \citep{1995ASPC...85....3W}.  Polars host strong $B\sim$7-230 MG magnetic fields and are readily detectable by their strong, soft X-ray emission \citep{1999hxra.conf..410B,2015SSRv..191..111F}.  The prototype for intermediate polar (IP) systems was DQ Her, and later, AE Aqr, which exhibit multiple optical and perhaps X-ray periods, although these pulsations are unpolarized, or only weakly polarized.  IPs are bright, hard X-ray sources \citep{2006MNRAS.372..224B}.  The strong magnetic moments in polars cause synchronous rotation with the binary orbital period.  IPs have weaker magnetic fields of $B\sim$1-10 MG, which do not lead to synchronous rotation, and as a result, the white dwarf primaries in these systems rotate more quickly than the system orbital periods.  The non-thermal  radio emission from magnetic CVs suggests that they may be divided into quiescent, weakly polarized, emitters of mildly relativistic synchrotron, or gyrosynchrotron, radiation and more powerful sources that exhibit highly circularly polarized radio emission driven by the electron cyclotron maser (ECM) \citep{2017AJ....154..252B}.
 
% \subsection{Two exceptional cases: AR~Sco and AE~Aqr - highly asynchronous polars with non-thermal emission}

There are  two exceptional IPs, AR~Sco and AE~Aqr, where the spin of the WD is extremely rapid compared with the  system orbital period. These extreme asynchronous polars have exceptional observational properties that have not been well explained from the
standard model of polars. Understanding the properties of these exceptional systems  is the main goal of the present work.

%The details of the interaction of the white dwarfs  in close binaries are also crucial for the further development of the theory of close binary evolution.

Most importantly, both  AR~Sco and AE~Aqr show high levels of non-thermal emission  extending  from radio to optical and X-rays.  AR Sco, dubbed   a ``white dwarf pulsar'',   shows modulations on the WD's spin and spin-orbital beat frequencies \citep[see][]{2016Natur.537..374M,2017NatAs...1E..29B,2017ApJ...835..150K,2018ApJ...853..106T,2018A&A...611A..66S,2019ApJ...872...67G}.  In contrast,   no isolated \WD\ produces pulsed radio emission \citep{2000PASP..112..873W,2017AJ....154..252B}.

%Gamma-ray emission section has been questioned and needs investigation

% Lamb \& Patterson (1983) suggested, based on spin-up arguments, that the magnetic field is ~ 105 G while a possible detection of circularly polarized radiation (Cropper 1986) implies a field in excess of 106 G.

%{\bf  AR Sco system}. 
AR Sco is arguably the most peculiar CV.
 On the one hand, the system appears similar to a CV system in that it hosts an M4 dwarf secondary that orbits a WD primary.  However, additional observed properties defy classification: (i) its spectral energy distribution (SED) is dominated by a modulated non-thermal component with power $L = 0.6-3.6 \times 10^{32} $ erg s$^{-1}$; (ii) The WD is spinning down at a rate $ \dot{P}=  4 \times 10^{-13}$; for a typical moment of inertia of a WD the corresponding spindown luminosity  $L_{sd}$ is few $ \times 10^{33}$ erg s$^{-1}$. This exceeds the emitted power by a factor $\sim 10$; 
% \item spin-down luminosity is about 11.5 times the mean electromagnetic power
(iii) There is bright variable optical emission at the beat of the WD and orbital periods;
(iv) Optical emission is highly linear polarized at 40\%, modulated both  on the
harmonic of the spin and beat period; (v) X-ray luminosity is fairly low, consistent mostly with thermal bremsstrahlung.  (Weak high energy emission excludes accretion as a driving mechanism.)
%\item The orbital phase of the maximum of optical emission is displaced from conjunction by about $0.15$ of the orbit.
(vi) WD mass is limited to $ 0.8 M _\odot  < M_{WD} < 1.29 M _\odot $;
(vii) There is variable high frequency ($\nu\sim 10$ GHz) radio emission from the RD that exhibits strong orbital modulation while the low frequency ($\nu\sim 1.5$ GHz) emission is relatively steady (importantly, there is no modulation in radio at the WD's spin period - this excludes the WD's \ms\ as the {\it locus} of radio emission).

 The properties of the system imply that  (i) the  spin-down time 
$
\tau_{sd} = \Omega/\dot{\Omega} \approx 10^7 {\rm  yr}$  is much smaller that the  age of the WD inferred from it's surface temperature $\sim 10^9$ yrs;
 (ii) the short spin period of the WD requires  a previous accretion stage to be spun up; (iii) low  X-ray luminosity excludes accretion as an energy source; (iv)
 the WD  light cylinder, which has a radius of $ 6 \times 10^{11}$  cm, is  $\sim 10$  times the orbital separation of the two stars.
(v) The RD is nearly Roche lobe-filling; it is also tidally locked.

 In this paper we  first concentrate on the   AR Sco system, and later on, \S \ref{AEAqr}, apply the results to  AE Aqr.

\section{Models of the torque on the WD that do not work}

Somewhat unconventionally, let 
us first provide a critique of the current models of AR Sco. First we will discuss what does not  work, and  later in \S  \ref{load} describe a   model  that  is able to explain the salient features of the system.
\subsection{Not a WD pulsar}
\label{notpulsar}

It is clear that the system involves interaction of the  stars'  \mss\ (or wind-\ms\ or wind-wind interaction). No isolated WDs come close to having the parameters  of AR Sco (\eg\ spin-down power).
In this sense, it is  different from radio pulsars, which produce  coherent radio emission in isolation.   No isolated WD produces radio emission, whether pulsed or steady \citep[\eg][]{2000PASP..112..873W}. 
In the   case of AR Sco (and AE Aqr) it is  clear that it is the interaction of the magnetosphere of the primary with that of the secondary that accelerates the emitting particles, even though ultimately it's the  rotational energy of the primary that gets converted into radiation. 
 The fact that only binary WDs, such as AR Sco and AE Aqr, produce radio emission may be explained by the necessary evolutionary channel: WDs need to be spun up by accretion in order to produce sufficient electric potential. Also, radio emission from AR Sco need not be coherent.

%MR: need to make sure that I link this last sentence to the radio section, or that we make sure to develop this argument elsewhere.

The biggest challenge in understanding the system, in our view, is to reconcile large present  spin-down rates and the requirement of previous spin-up of the WD.
Qualitatively, the large current  spin-down (seems)  to imply large \Bfs, while the need to previously spin-up  the WD requires small \Bfs. 
{\it The magnetic field on the WD should be low, as we argue next.} 

 First, the vacuum dipole formula for WD spin-down  is inapplicable: astrophysical plasmas always have enough charges available to screen parallel \Ef\ (only in rare localized circumstances like gaps in the \mss\ of  \NS\  do some mild $E_\parallel$ appears \citep{GoldreichJulian,Sturrock71,1977ApJ...217..227F}). 
This is especially true since the WD's  light cylinder radius is much larger than the separation  of the stars - the RD produces a dense wind
% {\bf (maybe include estimated density)}
that would make the vacuum approach for WD spin-down  invalid. 

Second, the possibility of a pulsar-like  spin-down also does not work for AR Sco.
Pulsar spin-down (though qualitatively similar to the vacuum  dipole case, but physically highly different)  was also suggested \citep[\eg][]{1998A&A...338..521I,2006A&A...445..305I,2008arXiv0809.1169I,2012ARep...56..595I}. The idea is  that the WD generates pair-dominated  pulsar-like wind \citep[hence the term ``White Dwarf Pulsar",][]{2017NatAs...1E..29B}.  Pulsars generate large electric potential drops along \Bf\ lines  that lead to vacuum breakdown, pair creations \citep{reesgunn,1977ApJ...217..227F}. These processes are accompanied by abundant $\gamma$-ray production.   Pulsars are bright $\gamma$-ray sources  \citep{2013ApJS..208...17A}.
It is possible that WDs can also  break vacuum \citep{1988SvAL...14..258U}. There is a clear prediction for this model: production of high energy emission that accompanies  pair production.  The available potential
 in AR Sco, $\Phi \sim \sqrt{L_{sd}/c}\sim 10^{14}$ eV matches the  weakest $\gamma$-ray pulsars. For example, one of the brightest $\gamma$-ray pulsar, Geminga, is located at 250 pc (about 2.5 times further than AR Sco)  and  has spin-down power of $3.2 \times 10^{34}$  erg s$^{-1}$  (about ten times higher). Although some $\gamma$-ray pulsars  do have  smaller spin-down powers than AR Sco \citep{2013ApJS..208...17A}, we disfavor this possibility, as no $\gamma$-ray emission is seen \citep{2019arXiv190800283K}, and the  X-ray emission is very weak \citep{2016ApJ...832...35L}.

In addition  to the  theoretical problems outlined above,  both the  vacuum dipole and pulsar spin-down formulae presented earlier yield extraordinary high \Bf\ estimates for a WD   \citep[\eg][]{2017ApJ...835..150K}:
  \be
  B_{WD} \approx \frac{ c^{3/2} \dot{\Omega}^{1/2} I_{WD}^{1/2}}{R_{WD}^3 \Omega_{WD}^{3/2}}= 4\times 10^8{\rm G}
  \label{B001}
  \ee
  {This is an exceptionally  high \Bf\  for a WD.}

A high \Bf\ on the WD is also inconsistent with the requirement that during the preceding accretion state, the WD was spun up.  
 Assuming that during the  high  accretion rate stage all of the mass lost by the secondary accretes onto the WD, and using the corotation condition at the edge of the \ms\, 
\ba &&
r_c =\frac{ ( G M_{WD})^{1/3}} {\Omega_{WD}^{2/3} } = 4\times 10^{9} {\rm cm}=0.05 a
\nn &&
 r_A^{(a)}= \frac{B_{WD}^{4/7} R_{WD}^{12/7}}{(2 G M_{WD})^{1/7} \dot{M}_{RD}^{2/7} }
\label{Rc}
\ea
($r_A^{(a)}$ is the \Alfven radius during spin-up stage)
 the needed accretion rate during the spin-up stage is,
 \be
 \dot{M} = 4\pi \frac{B_{WD}^2 R_{WD}^6 \Omega_{WD}^{7/3}}{ ( G M_{WD})^{5/3}}
 \label{mdotrec}
  \ee
 Using estimate (\ref{B001}) for the \Bf\ evaluates to $ 1.6 \times 10^{-2} M_\odot {\rm yr}^{-1}$ which is unrealistic by many orders of magnitude.

%Also, the  vacuum dipole/pulsar spin down formulae assume clean \ms\ with very little mass loading (so that the \Alfven velocity is of the order of the speed of light). This limit  is clearly  are not applicable  since the WD's  light cylinder is about ten times the separations between the stars.

\subsection{Not  WD's magnetosphere -  RD star interaction }
\label{Katz}

Stellar winds created by the outflow of plasma  along  the open magnetic field lines are ubiquitous among stars and stellar remnants such as WDs.  In the particular case of a WD-RD binary the winds from both stars are magnetically driven.
 Depending on the location of the critical  (\Alfven) points in the winds, one can identify several cases: (i) wind-wind interaction (both \Alfven points inside the Roche lobes), 
 (ii) \ms-wind interaction (one \Alfven point is outside  the Roche lobe); (iii) direct magnetospheric interactions (both \Alfven points beyond the Roche lobe); (iv) if one of the winds is very weak, one can also envision direct 
 wind-star and \ms\ interactions.

%The RD in AR Sco has a radius $R_{RD} \sim 2.5 \times 10^{10}$ cm. At a separation of $8 \times 10^{10}$ cm the  M dwarf occupies $\sim (R_{RD}/a)^2 /4 = 0.02$ of the sky. But about 10\% of the spin luminosity has to be converted into radiation. Hence the interaction region should be larger than the size of the RD. 

    For a radius of $R_{RD} \sim 2.5 \times 10^{10} cm$, the ratio of the RD's radius to the separation is $R_{RD}/a=0.31$. For  the mass ratio $q\approx 0.4$  (the emission measurements lead to the  limit of  $q > 0.35$, Marsh)
and using 
 \be
\frac{R_{Roche}} {a}= \frac{0.49 q^{2/3}}{0.6 q^{2/3}+\log \left(\sqrt[3]{q}+1\right)}
 \ee
  for the size of the Roche lobe with $q=0.4$ \citep{1983ApJ...268..368E}, the size of the RD's Roche lobe is similar to it's radius -  the RD is nearly Roche lobe-filling.

 \cite{2017ApJ...835..150K} suggest that the interaction between the corotating WD \ms\ and the RD leads to higher spin-down rate of the WD.  
 %MR: Is this a comparison between the interaction of the WD magnetosphere with a wind-less, magnetized RD?
 On basic grounds, if a star with surface \Bf\ $B_{WD}$ and angular velocity $\Omega_{WD}$ interacts with a particularly resistive object of size $R_{int}$ located at distance $d_{int}$, the spin-down power can be estimated as 
 \be
 L_{sd} \approx \frac{B_{WD}^2}{4\pi} \left( \frac{R_{{WD}}}{d_{int}} \right)^6 \pi  R_{int}^2 d_{int} \Omega_{WD} 
 \label{Lsd0}
 \ee
 (this is a  magnetic  stress, assuming that the tangential component of the \Bf\ is of the order of the normal, times the interaction area, times the velocity of the field lines).
 % {\bf (Do you want to insert an equation here instead?)}
 
 If interaction is with the RD, so that $d_{int} \approx a$, $R_{int}= R_{RD}$, then the required \Bf\ is 

\be
B_{WD}  \approx 2 \frac{ a^{5/2} \dot{\Omega}^{1/2} I_{WD}^{1/2}}{R_{RD} R_{WD}^3 }= 4.4 \times 10^7{\rm G}
\ee
The  corresponding required accretion rate during spin-up (\ref{mdotrec}) is still unrealistically high, $\dot{M} \approx  10^{-4} M_\odot$ yr$^{-1}$.

\subsection{Not  WD's magnetosphere -  RD's magnetosphere  interaction}
\label{recon1}

\begin{figure}
\centering
\includegraphics[width=.99\textwidth]{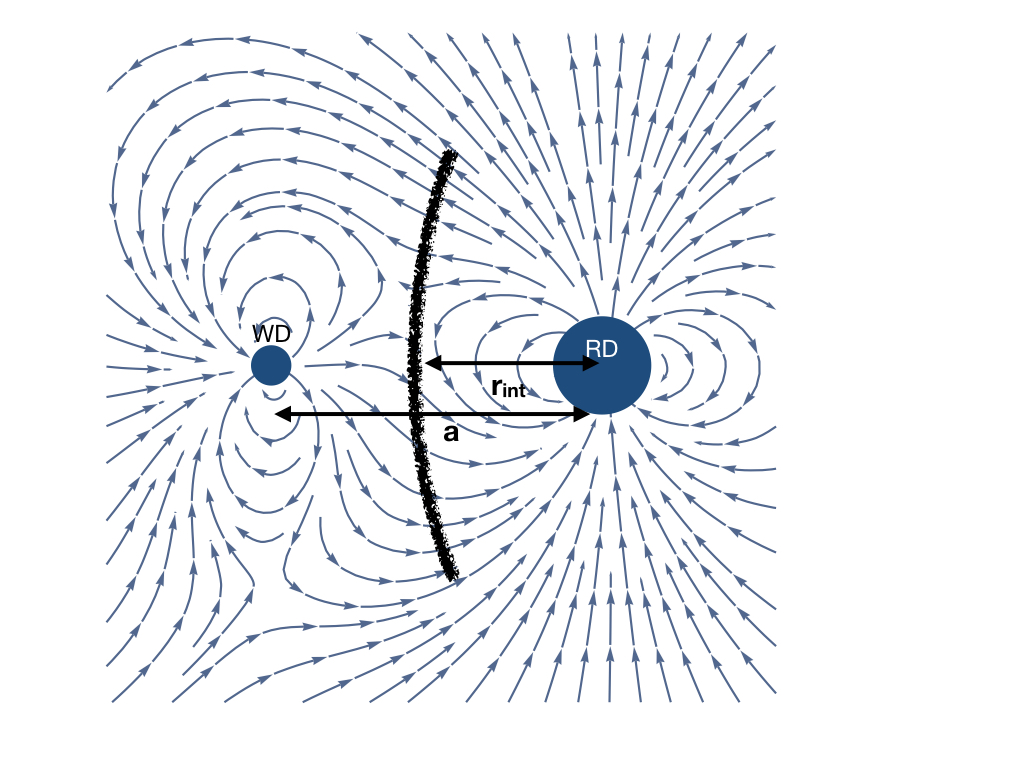}
\caption{Cartoon of direct magnetospheric interaction. Pictured is a poloidal slice (in the plane containing the orbital momentum and the line connecting two stars) of the interacting \Bf\   of the WD and the RD. The hashed line indicates the location of the reconnection region, where field lines connect the surfaces of two stars.  Orbital plane is horizontal, the magnetic moment of the WD is in the orbital plane, pointing at the moment towards the RD; the RD's magnetic moment is along the orbits' normal (pointing up); in the picture the magnetic moment of the RD is 5 times that of the WD. The structure of the \ms\ is north-south asymmetric: in half a period of the WD's rotation the asymmetry will reverse.   Note: object sizes and distances are not to scale. 
}
\label{WD-RD-int} 
\end{figure}

One possibility to increase the interaction size is through magnetic interaction of the two \mss\  or interaction of the WD's \ms\ with the extended wind of the RD.
In order to affect the WD spin-down the balance between the interacting WD's and RD's  flows should  be inside the WD's Alfven radius. The interaction is either between the solidly rotating WD's magnetosphere and the   RD's wind, or directly between two  magnetospheres. Here we discuss a case of direct magnetospheric interaction, Fig. \ref{WD-RD-int}. As we discuss below, the magnetically  interacting \mss\ cannot explain the WD's spin-down. Yet, this process is important for the generation of  emission, \S \ref{emit}.

% case  naturally explains many properties of the system -  this is our preferred model, Fig. \ref{WD-RD}. The case of wind-\ms\ interaction is considered in   Appendix \ref{load}.

Assume that the stars have surface \Bfs\ $B_{WD}$ and $B_{RD}$. For the given radii, $R_{WD} $ and $R_{RD}$ the force balance between two \mss\ occurs at distance  $r_{int}$ from the RD  given by
\ba &&
\frac{r_{int}}{a} = \frac{1}{1 +(R_{WD}/R_{RD}) (B_{WD}/B_{RD})^{1/3}} =  \frac{1}{1 +\eta_R \eta_B^{1/3}} =  \frac{1}{1 +\eta_\mu}
\nn &&
\eta _ B = B_{WD}/B_{RD} \gg 1
\nn &&  
\eta_R = R_{WD}/R_{RD} =0.02 \ll1 
\nn &&
\eta_\mu = \frac{\mu_{WD}}{\mu_{RD}}
\ea
  M dwarfs can have surface  \Bfs\ $\sim 10^3$ G; as a result, the RD's \ms\ can   extend beyond its Roche lobe. For a WD with surface \Bf\ of $10^6$ G the balance between the \Bfs\ will be at a distance $\sim 6.7 \times 10^{10}$ cm - way inside the WD's Roche lobe. 
  % (see Fig. \ref{rb}). 
  %Only for very low surface \Bf\ on the M dwarf, as little as  $1G$, the interaction of the WD's \ms\ will be  directly with the surface of the M dwarf. 
Only for a very low \Bf\ of the RD and extremely high \Bf\ of the WD, so that $ B_{WD}/B_{RD} \geq 10^6$, will the balance between magnetic pressures be inside the RD's Roche lobe. Thus, the interaction between \mss\ of the companions will generally be within the WD's Roche lobe.

At the balance point, the local \Bf\ is 
\be
\frac{B_{int}}{B_{WD}}= \frac{(1+\eta_\mu^{1/3} )^3}{\eta_B} 
\left( \frac{R_{RD}}{a} \right)^3 \approx
\left( \frac{R_{RD}}{a} \right)^3 \times 
\left\{
\begin{array}{cc}
\frac{1}{\eta_B}, & \eta_B^{1/3} \eta_R \ll 1
\\
 \eta_R^3  \left( \frac{R_{RD}}{a} \right)^3, & \eta_B^{1/3} \eta_R \gg 1
 \end{array}
 \right.
 \label{Bintt}
\ee
 In the particular case of AR Sco this requires $\eta_B \geq 10^5$; thus we expect $\eta_B^{1/3} \eta_R< 1$. In this regime the \Bf\ in the interaction region is independent of the \Bf\ of the WD:
 
 \be
 B_{int} \approx B_{RD} \left( \frac{R_{RD} }{a} \right)^3 
 \ee
 
 Using (\ref{Bintt}) as an estimate of the \Bf\ at the interaction region, $d_{int} \approx a - r_{int}$ (recall that $r_{int}$ is measured from the RD, and interaction size $R_{int} \approx a -r _{int}$, the spin-down power (\ref{Lsd0}) becomes
 \be
 L_{sd} = \frac{1}{4} \frac{(1+ \eta_\mu^{1/3})^3}{\eta_\mu} \frac{B_{WD}^2 R_{WD}^{6} \Omega_{WD}}{a^3}
 \label{Lsd00}
 \ee
 where we expressed all the quantities in terms of WD's parameters and the radial and magnetic ratios $\eta_R$ and $\eta_B$

In our case 
\be
\eta_\mu \equiv \left( \frac{R_{WD}}{ R_{RD}} \right)^3   \frac{B_{WD}}{ B_{RD}} = 8 \times 10^{-6}  \frac{B_{WD}}{ B_{RD}} 
\ee
For $B_{RD} \sim 10^3$ G it is likely to be much smaller than unity: $\eta_\mu \ll 1$. In this case (\ref{Lsd00}) gives

\be
 L_{sd} = \frac{1}{4}  \frac{B_{RD} B_{WD}  R_{RD}^3 R_{WD}^{3} \Omega_{WD}}{a^3 } =  \frac{1}{4} \frac{\mu_{WD} \mu_{RD}} {a^3} \Omega_{WD}
 \label{Lsd11}
 \ee
 The required \Bf\ is then
 \be
 B_{WD} = 4 \frac{a^3 \dot{\Omega} I_{WD} }{ B_{RD} R_{RD}^3 R_{WD}^3} =  2 \times 10^7 {\rm G}
 \ee
 The necessary $\dot{M}$, Eq. (\ref{mdotrec}),  is still too high, $\dot{M} \sim 3\times 10^{-5} M_\odot $ yr$^{-1}$. 

Thus we conclude that magnetospheric interaction, the most efficient of the scenarios considered, cannot accommodate the requirements of large current spin-down and efficient spin-up during the accretion stage.    Importantly,  the magnetically  interacting \mss\ cannot explain the WD's spin-down, \S \ref{recon1}, yet this process is important for the generation of emission, as we will describe further in \S \ref{emit}.

Below, in \S \ref{load}, we demonstrate that the WD's spin-down can be easily explained due to mass loading of the RD's wind onto the corotating WD's \ms.

\section{The model of the WD's torque:  mass loading  from the RD}
\label{load}
\begin{figure}[h!]
\centering
\includegraphics[width=.99\textwidth]{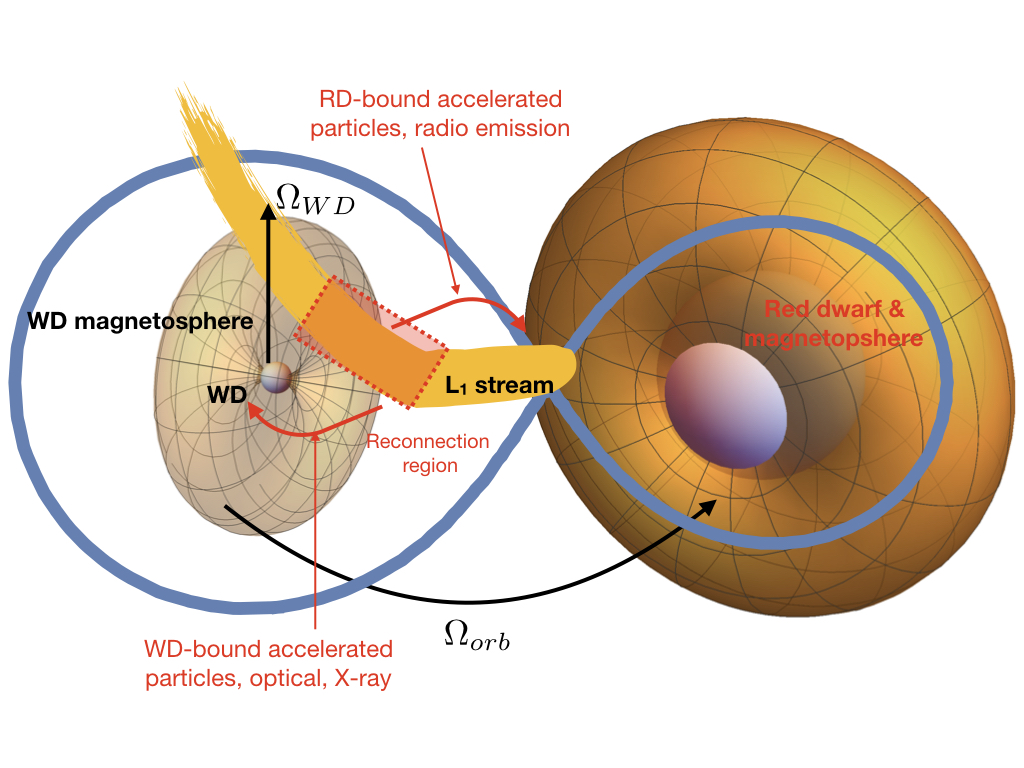}
\caption{Cartoon of the model. The corotating  \ms\ of the WD extends beyond the WD's Roche lobe and interacts with the wind/\ms\ of the red dwarf. The red dwarf loses mass through the $L_1$ point. The partially ionized accretion stream loads the \ms\ of the WD, providing large torque on the WD as it is ejected in the propeller regime.
 Previously, when the mass loss rate from the RD was high, the WD was spun-up in the accretor regime. Nonthermal particles are accelerated in the reconnection region between the RD's accretion flow and the WD's \ms, producing optical and X-ray emission near the WD and radio emission near the RD. }
\label{WD-RD} 
\end{figure}

 In \S \ref{notpulsar} we demonstrated that arguments in favor of high \Bfs\ are untenable. 
We concluded  then that  the WD's  spin-down is due to the interaction with a companion.  The key point then is  to understand the WD-RD  interaction and how it affects the WD spin-down and production of radiation. 
In this Section we  discuss a model that can satisfy both the condition of large current spin-down, and the requirement of  the low \Bf\ from the spin-up conditions, as shown in Fig. \ref{WD-RD}.

\subsection{Magnetic field of the WD must be low}

We can estimate the WD's \Bf\ using the condition that for a given  mass loss rate from the RD, $\dot{M}_{RD}$, accretion onto the WD spins up the latter. Using (\ref{mdotrec}) with $r_c= r_A^{(a)}$ we find
\be
B_{WD} = \frac{ \sqrt{2}  \dot{M}_{RD}^{1/2}  ( G M_{WD})^{5/6}}{  R_{WD}^3 \Omega_{WD}^{7/6}}
=5\times 10^5 {\rm G}
\label{B5}
\ee
for the maximal  accretion rate of  $ \dot{M}_{RD, max} \sim 10^{-9}M_\odot$   yr$^{-1}$ \citep{1981A&A...100L...7V,2011ApJS..194...28K}. 
 Thus, the WD's \Bf\ {\it must be sufficiently low to allow spin-up}.

\subsection{Mass loading from  spherical  RD wind?} 
\label{sec:spherical}
  
Let us first discuss a toy model with  mass loading from a spherical RD wind. This simple approach will allow us to make estimates of the main parameters of the system. As we discuss below, \S \ref{sec:straem}, the actual mass loading occurs via Roche lobe overflow.

Assume that mass loading of the WD's \ms\ occurs at radius $r_A \approx v_A/\Omega$ with rate $\dot{M}_{WD}$. In the propeller regime (the current state) the loaded material is ejected with velocity $\sim r_A \Omega_{WD}$. The  system is governed by the following set of conditions
\ba &&
 L_{sd} = \dot{M}_{WD} (r_A \Omega_{WD})^2
 \nn &&
 v_A= \frac{B}{\sqrt{4\pi \rho}}
 \nn && 
 B=B_{WD} \left( \frac{r_A}{R_{WD}}\right)^{-3}
 \label{eq:Lwd}
 \ea
 and
 \ba &&
 \dot{M}_{WD}  = 4\pi \rho v_A r_A^2
 \nn &&
  \dot{M}_{WD}= \frac{r_A^2}{4 a^2} \dot{M}_{RD}
  \label{eq:mdotan}
  \ea
  where we assumed that the relative fraction of the mass loaded onto WD's \ms\ is proportional to the mass loss rate of the RD $ \dot{M}_{RD}$ and the relative fraction of the RD's sky occupied by the interaction region, $\sim {r_A^2}/({4 a^2})$.
  
  Equation (\ref{eq:mdotan}) 
  gives
  \ba &&
  r_A = \frac{ \sqrt{2 a} L_{sd}^{1/4}}{\dot{M}_{RD}^{1/4}  \Omega_{WD}^{1/2}} = 2 \times 10^{10} {\rm cm}
  \nn &&
  B_{WD} = \frac{ (2 a)^{3/4}  L_{sd}^{7/4}}{\dot{M}_{RD}^{3/8}  R_{WD}^3 \Omega_{WD}^{5/4}} = 3\times 10^6 {\rm G}
  \nn &&
   \dot{M}_{WD}= \frac{ \sqrt{\dot{M}_{RD}  L_{sd} }}{2 a \Omega_{WD}}= 1.4 \times 10^{-11} M_\odot {\rm yr}^{-1}
   \label{eq:mdotnum}
  \ea
   for $\dot{M}_{RD}= 10^{-9} M_\odot {\rm yr}^{-1}$. 
   
   Thus, in order to account for the  large spindown of the WD, the spherical accretion requires a very large mass loss rate from a RD. 
   In the following, we develop model of WD's mass loading through Roche lobe overflow and ensuing ionization.  

\subsection{Mass transfer via Roche lobe overflow}

The atmospheres of RDs are relatively cold and dense; they are expected to be partially ionized. If the neutral-ion collision rate in the RD's wind is not high 
\citep[this is far from certain; see][]{2019ApJ...872...67G}, then the neutrals from the RD wind will stream freely onto the \Bf\ lines of the WD. 
They will be exposed to the UV radiation from the surface of the WD that will lead to ionization. As the neutrals get ionized they will couple to the \Bf\ of the WD, 
and will be centrifugally  expelled from the system. Below we give order-of-magnitude estimates for the efficiency of ionization, leaving a more detailed analysis to a subsequent paper. 

Next we discuss the RD-WD interaction that explains the key features of the WD's spin-down due to loading of the WD's \ms\ by the partially ionized RD' wind.
We envision two possible scenarios: accretion onto the WD from a spherical wind from RD, \S \ref{sec:spherical}, and   accretion via a tightly confined matter stream caused by the Roche lobe overflow, \S \ref{sec:straem}.

\subsubsection{The temperature of the WD}
\label{TWD}
The ionization efficiency of the WD's radiation depends sensitively on its surface temperature.   \cite{2016Natur.537..374M}  reported a surface temperature of $T_{WD} = 9750$~K, although it may be as high as $T_{WD} \approx  12000$~K, as we describe below based on the analysis of \emph{Hubble Space Telescope (HST)} data. This difference has important implications for the ionization processes in the wind, \S \ref{processes}.

\begin{figure}[h!]
\centering
\includegraphics[width=.9\textwidth]{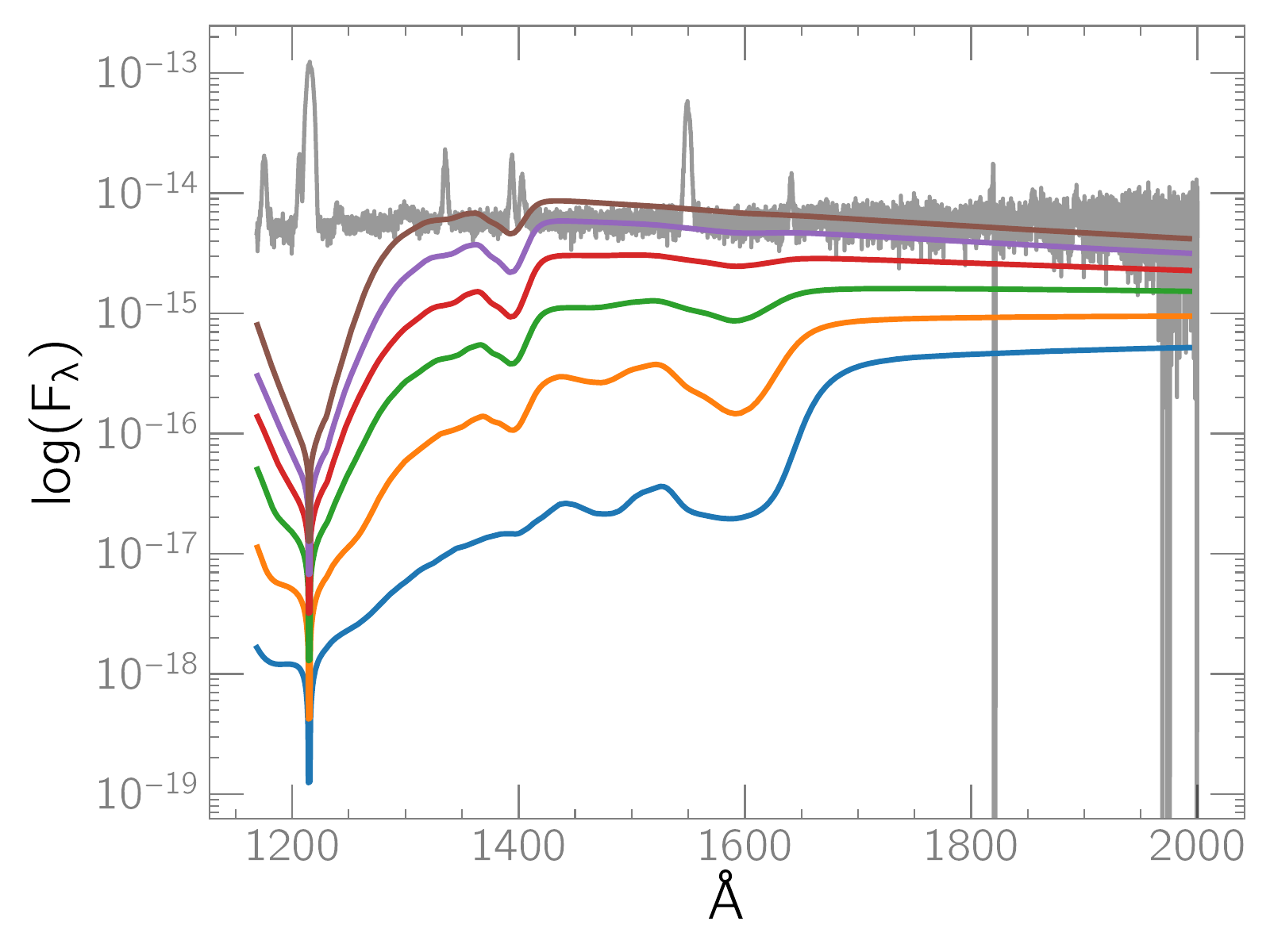}
\caption{A comparison between the averaged \emph{HST}/COS spectrum of AR~Sco (grey line) and  \protect\cite{2010MmSAI..81..921K}  WD models scaled to the Gaia distance. The observed spectrum is a sum of the WD contribution, the varying synchrotron radiation from the interaction, and line emission from the irradiated face of the secondary. 
%The Lyman-$\alpha$ line is dominated by the geocoronal emission.
The six colored lines represent WD models for log(g) = 8.5 and differ only in their effective temperatures.
The temperatures range from 10,000 K (blue line) to 15,000 K (brown line) in increments of 1,000 K.  Models with temperatures
higher than 13,000 K predict that the photospheric contribution of the WD would be detectable, establishing an upper limit of $T_{WD}\leq 13000$~K for the WD temperature. As noted in the text, the  \protect\cite{2010MmSAI..81..921K} spectra neglect magnetic effects.}
\label{temperature} 
\end{figure}

To constrain the WD's effective temperature, we downloaded a grid of the \cite{2010MmSAI..81..921K}  WD atmospheric models spanning a wide range of effective temperatures. We assumed a surface gravity of log(g) = 8.5. We then scaled the spectra to the Gaia distance of AR Sco (d=117 pc), assuming a WD radius of 7,000 km. Finally, we plotted the scaled spectra and compared them against the \emph{HST}/Cosmic Origins Spectrograph(COS) spectrum \citep{2016Natur.537..374M},  Fig. \ref{temperature}.. We found that for $T_{WD}\gtrsim 13,000$~K, the WD's photospheric contribution would be detectable in the \emph{HST} spectrum, so we adopt this as a upper limit for the WD's effective temperature.

A  limitation of this approach is that the \cite{2010MmSAI..81..921K} models neglect magnetic effects. Given the unknown magnetic field strength of the WD, Zeeman splitting could have a significant impact on the WD's photospheric lines. Higher signal-to-noise ratio UV spectra obtained around orbital phases when the system its faintest would provide more stringent limits on the WD properties.
% {\bf (Should we change this to be something like ``during primary eclipse''?)}.

\subsubsection{Ionization of the RD's  stream}
\label{processes}

For a WD surface temperature $T_{WD} \le 12000 $~K, \S \ref{TWD},  the number of photons emitted above the ionization threshold $\nu_0$ is
\be
\dot{N}_{ph} = 4 \pi R_{WD}^2 \int_{\nu_0}^\infty 4\pi \frac{\nu^2}{c^2}  \left( e^{h \nu/T} -1\right)^{-1} d \nu = 5.5 \times 10^{38} {\rm s} ^{-1}
\label{Nph}
\ee
where $\nu_0=3.29\times 10^{15}$~Hz is the frequency corresponding to Hydrogen ionization.
The corresponding mass loading rate for complete absorption would be  
$\dot{M}_{load} =m_p \dot{N}_{ph}= 1.4 \times 10^{-11} M_\odot   {\rm  yr} ^{-1}$. 

Also, the effective optical depth for ionization is of the order of unity
\ba &&
\sigma_i  = \sigma_0 (\nu/\nu_0)^{-3}
\nn &&
n_n = \frac{\dot{M}_{RD} }{4\pi   m_p a^2 v_{w,RD}}
\ea
where $n_n$ is the density of neutrals, scaled to the RD mass loss rate of $10^{-9} M_\odot$ yr$^{-1}$, $\sigma_i$ 
is the ionization cross-section, in units of $ \sigma_0 = 6.3 \times 10^{-18}$ cm$^2$. Also, recombination time scales are mostly likely long enough.
So the optical depth of the system for ionizing radiation is
\be
\tau(\nu) \approx n_n \sigma_i a  \approx 8000\frac{\dot{M}_{\rm RD,-9}}{a_{10.9}^2 v_{w,7.5}}
\left(\frac{\nu_0}{\nu}\right)^3.
\label{eq:tau}
\ee

The direct ionization radius can be estimated from equation
\be
\frac{R_{\rm WD}^2}{r_{\rm io}^2}\int_{\nu_0}^\infty d\nu B_\nu \sigma_i=\frac{\max(v_{\rm w},v_{\rm ff} ) }{r_{\rm io}}
\label{eq:rio}
\ee 
where  $B_\nu$ is Planck's spectrum  
%{\bf use $B_\nu$ in \ref{Nph} }  {\bf (move closer to Eq. 21?)},
\be
v_{\rm ff} = \sqrt{\frac{G M_{\rm WD}}{r_{\rm io}}}
\label{eq:vff}
\ee
even for $T_{\rm WD} = 9750$~K and stellar wind with speed about 300~km/s the ionization radius evaluates to $r_{\rm io}\approx 5\times 10^{11}$~cm; this  exceeds the  orbital separation by  more than 5 times.

In the previous estimation we neglect the absorption of ionizing photons by neutrals. Nevertheless, the secondary photons can ionize the matter in the stellar 
wind, the  mass flux of neutrals can be balanced by ionizing photons flux or $\dot{N}_{ph} = \pi r_{io, max}^2 v_w n_n$. 
\be
r_{\rm io,max} \approx \sqrt{\frac{\dot{N}_{ph}}{\pi n_n v_{\rm w,RD}}} \approx 1.9\times 10^{10} \dot{M}_{RD,-9}^{-1/2} a_{10.9} \mbox{ cm}
\label{eq:riomax}
\ee
This corresponds to WD magnetosphere injection rate on the level
\be
\dot{M}_{WD} \approx \frac{r_{\rm io,max}^2}{4a^2} \dot{M}_{\rm RD} \approx \dot{N}_{ph} m_p \approx 1.4\times 10^{-11} M_\odot /\mbox{yr}
\label{eq:mdotinj}
\ee 
So, if $r_A\ge r_{\rm io,max}$ in the spherically symmetric case the WD magnetosphere loading rate  depends on ionizing photon production rate only.

Thus, we can estimate the mass loading rate  using two different methods: Eq.~(\ref{eq:mdotnum})  and Eq.~(\ref{eq:mdotinj}). The required WD's temperature is then $T_{WD} = 12000 $~K. 
In contrast,  \cite{2016Natur.537..374M} estimated  $T_{WD} \le 9750 $~K; this supplies  $\sim 1/30$ of the required UV photon production rate.  A new series of observations by \emph{HST} at ``off the peak''  orbital phases
could clarify this problem.

In conclusion,  we expect that the WD's radiation can ionize hydrogen in the outer parts of the RD's corona, and in  the surrounding area. On the other hand, if 
the mass flow from the RD is large, it can screen the ionizing radiation, so that the neutral component of the RD's flow can penetrate the WD's \ms.

\subsubsection{Mass transfer rates via Roche lobe overflow} 
\label{sec:straem}

Next  we discuss  a more realistic scenario based on   mass transfer via Roche lobe overflow. In this case the main mass transfer process takes place through the first Lagrangian 
point $L_1$. As a result, the wind mass loss rate  from the RD can be much smaller than for the spherical wind case discussed in the previous subsection \S \ref{sec:spherical}. As a result, for smaller ``effective'' (isotropic-equivalent) mass loss rates, the   radiation from the WD can 
ionize the  photosphere of RD. Still the   matter flowing through the  $L_1$ point can contain neutral components.

In this  case  Eqns.~(\ref{eq:Lwd})  and (\ref{eq:mdotan}) give 
 \ba &&
 \dot{M}_{WD}  \approx \eta_s \rho_s v_A r_A^2
 \nn &&
  \dot{M}_{WD} \approx \xi \dot{M}_{RD, stream}
  \label{eq:mdotanstr}
  \ea
here $\rho_s$ is density of the stream, $\eta_s$ is a constant order of 1 which take into account the geometry factor of the stream, $\xi$ is a fraction of the steam ionized and accelerated in WD magnetosphere.  
 \ba &&
  r_A = \left(\frac{L_{sd}}{\xi\dot{M}_{RD, stream}\Omega_{WD}^2}\right)^{1/2}= 2 \times 10^{10} \, {\rm cm}
 \nn &&
  B_{WD} = \left(\frac{4\pi L_{sd} r_A^3}{ \eta_s \Omega_{WD} R_{WD}^6}\right)^{1/2} = 1.3 \times 10^7 \eta_s^{-1/2} {\rm G}
  \label{eq:mdotanstr1}
  \ea
where   $\xi\dot{M}_{RD, stream} \sim 10^{-11} M_\odot {\rm yr}^{-1}$ was assumed.

%\textbf{Maxim L.: Synchrotron UV emission also  can ionize hydrogen. Can we restrict $\dot{M}_{\rm RD}$ using observed flux in UV? It can be self regulating process...}

The $r_A$ and correspondingly $v_A$ are the same in Eq.~(\ref{eq:mdotanstr1}) and Eq.~(\ref{eq:mdotnum}), so the flow rate should be the same.  Following the analysis in \S \ref{sec:spherical} we can estimate the magnetosphere mass loading rate due to  ionization of neutrals  flowing  through the $L_1$ point and the ionizing photons number as  
\be
\dot{M}_{WD} \approx \frac{\dot{N}_{ph} m_p}{\eta_s} \approx 1.4\times 10^{-11} M_\odot /\mbox{yr}.
\label{eq:mdotinjs}
\ee
In the case of the flow from $L_1$ point, the ionization photons flux should be in $1/\eta_s$  if compared to  the spherical wind case.

%{\bf (I don't think I understand.  This is the same result as Eq. 17, which had the denser RD wind.  Are we using the degree of photoionization as a tunable parameter that makes the two calculations the same?)}

 The source of the UV photons can be both the  WD as well as the nonthermal synchrotron radiation from the interaction region. We hypothesize that in the latter case, a self-regulating quasi-periodic system evolves through the following ionization states:
 %with  period in between WD spin and fraction of orbital period. The cycle can be following:
  1) ``Plunging'':  no nonthermal emission: the stream goes deeply into magnetosphere where it starts to be ionized; 2) ``photoionization'': strong interactions between the ionized matter stream and 
magnetosphere produce nonthermal radiation which starts to ionize matter in the stream; and 3) ''quenching'':  the strongly ionized stream stops penetrating and interacting with the WD magnetosphere, leading to suppression of nonthermal emission 
and returning the system to phase 1. So, the system will oscillate around the equilibrium state.% This explains the close values of direct ionization radius and total ionization radius and Alfvenic radius. REFERENCE?? AFTER EQ. 26,  AND (23)

\section{Emission model: acceleration at reconnection between interacting magnetospheres}
\label{emit}

\subsection{Acceleration at reconnection}
Interaction of the \Bfs\  between  the WD's and the RD's \mss\ will lead to reconnection. 
Particles will be heated and accelerated in the reconnection events. The  reconnecting \Bfs\ connect back to the WD and to the RD, where the accelerated particles will produce synchrotron/cyclotron emission within the corresponding  \mss. 
The synchrotron origin of optical emission at the WD is consistent with the highly linearly polarized optical signal, showing the polarization rotation \citep{2017NatAs...1E..29B,2019arXiv191007401D}, similar to the   rotating vector model in pulsars \citep{1969ApL.....3..225R}, see \S  \ref{optical}. The cyclotron origin of the radio emission in the RD is discussed in \S \ref{radio}.

The reconnection events are expected to produce signal at the beat frequency between the WD's spin and the orbital motion: as the  field lines from different magnetic poles of the WD sweep by the MD, the polarity of the  \Bf\ in the wind changes every half a period. Depending on the orientation of the \Bf\ of the MD, reconnection between the wind and MD's \Bf\ occurs every period.

The reconnection between the WD's wind and MD's \ms\ should proceed in a somewhat typical fashion. The two plasma components have  different plasma properties: very light WD's \ms\ and relatively heavy RD's \ms. Hence we expect different properties (density and temperature) on the two sides of any reconnection point. 
 \begin{figure}
\centering
\includegraphics[width=.99\textwidth]{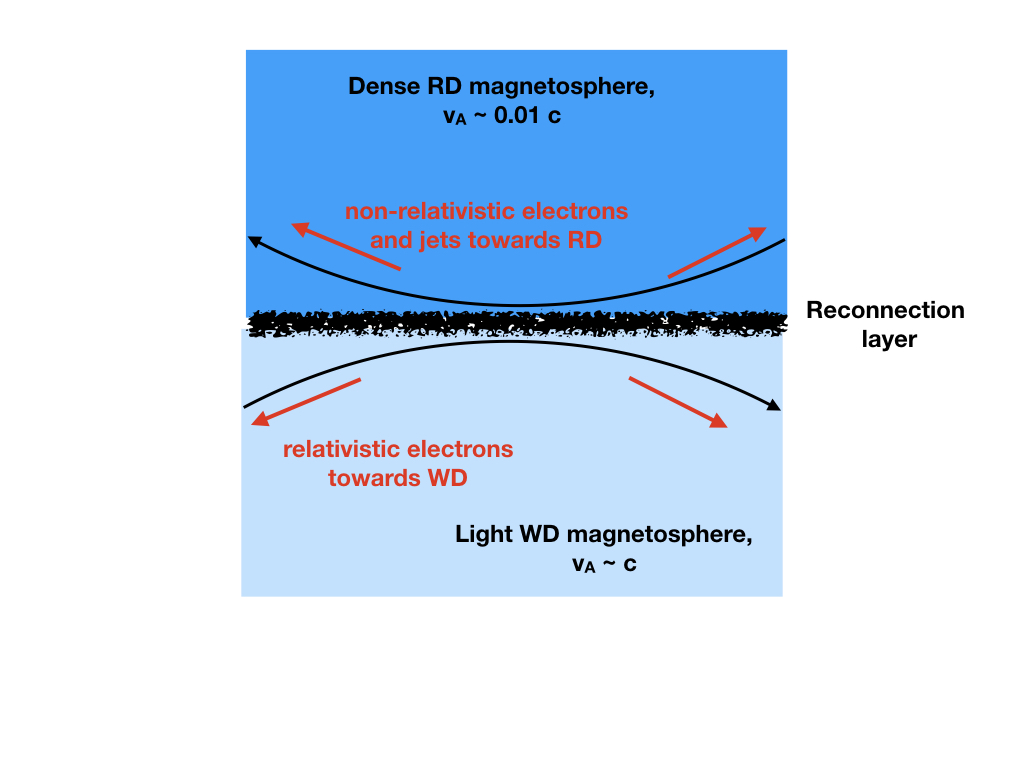}
\caption{Asymmetric reconnection layer between the dense RD \ms\ and the rarefied WD \ms.}
\label{WD-RD-rec} 
\end{figure}

%On the WD side, the plasma beta, $\beta = 8 \pi p /B_{int}^2 \ll 1 $, is very low. In fact, plasma sigma, $\sigma = B_{int}^2/ (4 \pi \rho)$ can be of the order of unity. (If the solid body rotation extends to the light cylinder, then the required  $\sigma$  satisfies:  $v_A = \sqrt{\sigma/(1+\sigma)}$, hence $\sigma \geq ( c/(a \Omega_{WD}) )^2 \approx 10^{-2}$.) Reconnection in such plasmas proceeds in  specific, unusual (from the classical point of view)  regimes \citep[\eg][]{2017JPlPh..83f6301L,2017JPlPh..83f6302L}. Particles can be, under very extreme conditions, quickly  accelerated up to the maximal available  potential. The maximal \Lf\ of particle can be then estimated as a potential across the reconnection region of size $\sim a$, \Bf\ $B_{int}$  and velocity of incoming \Bf\ lines $\Omega a$ (so that \Ef\ $E \sim ({\Omega a }/{c}) B_{int}$)
%\ba &&
%\gamma_{max} \approx \frac{e \Phi}{m_e c^2} = \frac{e}{m_e c^3} \frac{B_{WD} R_{WD} \Omega}{a} = 10^6
%\nn &&
% \Phi \approx a \frac{\Omega a }{c} B_{int}
% \label{gammamax}
% \ea
% This is a very high \Lf, but this is the upper estimate. As we demonstrate below, Eq. (\ref{em}),  the  \Lf\  of the electrons accelerated towards the WD need to be $\sim 10^{-3}$ of the maximal possible value.

In the frame of our model we expect strong deformation of the WD's magnetosphere due to interaction with stellar wind at the Alfvenic radius. Therefore,  the plasma beta, $\beta = 8 \pi p /B_{int}^2 \approx 1 $.  Reconnection in such plasmas proceeds in  specific, unusual (from the classical point of view)  regimes \citep[\eg][]{2017JPlPh..83f6301L,2017JPlPh..83f6302L}. Particles can be, under very extreme conditions, quickly  accelerated up to the maximal available  potential. The maximal \Lf\ of particles can be then estimated as a potential across the reconnection region of size $\sim r_A$, \Bf\ $B_{int}=B_{\rm WD}(r_A/R_{\rm WD})^{-3}\sim 200$~G  and velocity of incoming \Bf\ lines $\Omega r_A$ (so that \Ef\ $E \sim ({\Omega r_A }/{c}) B_{int}$)
\ba &&
\gamma_{max} \approx \frac{e \Phi}{m_e c^2}  \sim 10^7
\nn &&
 \Phi \approx r_A \frac{\Omega r_A }{c} B_{int}
 \label{gammamax}
 \ea
 here we substitute values from Eq.~\ref{eq:mdotanstr1}.
 This is a very high \Lf, but this is the upper estimate. As we demonstrate below, Eq. (\ref{em}),  the  \Lf\  of the electrons accelerated towards the WD need to be $\sim 10^{-3}$ of the maximum possible value.

On the RD side the reconnection will be analogous to the solar \ms, where particles are heated and produce UV and soft X-ray emission; this explains the X-ray emission from AR Sco. Non-relativistic exhaust jets that propagate with the local \Alfven velocity  couple to the neutral component in the MD atmosphere/corona and generate H$\alpha$ features  observed by \cite{2019ApJ...872...67G}. Particles are also accelerated to mildly relativistic energies and produce radio emission both in the interaction region and within the RD corona.

\subsection{Optical and  X-ray emission: \ms\ of the WD}
\label{optical}

We expect that in reconnection events particles are accelerated to power-law distributions. In the highly variable \Bf\ of the WD's \ms\ particles accelerated in the interaction region will be propagating downward, increasing their emitted synchrotron  frequency, while   losing energy to synchrotron emission (and reflected due to magnetic bottling). It is a fairly complicated problem how to calculate synchrotron emission from a stream of particles propagating with the \mss:  (i) the basic cyclotron frequency $\om_B$ changes with radius; (ii) in a collision-less plasma the particles' pitch angles change with radius due to conservation of the first adiabatic invariant; (iii)  the number of particles that reach a given radius changes due to the bottling effect; (iv) particle pitch angles evolves due to radiative losses \cite[\eg][]{2005ApJ...634.1223L}. We leave a more detailed consideration to subsequent paper. Here we just provide order--of-magnitude estimates.  In what follows we employ a concept that  starting from the emission region with a pre-defined \Bf, for a given emission frequency and radiated power there are optimal parameters to produce  emission subject to the above constraints.

As  an order-of-magnitude estimate, we assume that emission is dominated by particles with the synchrotron cooling time of the order of $c/r_{em}$ (lower energy particles do no emit efficiently since power $\propto \gamma^2$, while higher energy particles do not probe high \Bfs). To produce synchrotron emission at a frequency $\om$ and overall  power $L_s\sim  10^{32} $ erg s$^{-1}$ we need the number of particles $N_p$ emitting typically at distance $r_{em} $ to be such that:
\ba &&
\omega \approx \gamma _ {em}^2 \om_B 
\nn &&
\tau_ c \approx \frac{ m_e c^3}{e^2 \gamma _ {em}\om_B^2 }= \frac{r_{em}}{c}
\nn &&
L_s \approx N_p \frac{e^2}{c} \gamma_ {em} ^2 \om_B^2 
\label{44}
\ea
The above relations apply to  particles  in both \mss,  
\be
B = B_{{WD} } \left( \frac{r_{em}}{R_{WD}} \right)^{-3}
\label{441}
\ee
where $ B_{{WD} } $  and $R_{WD}$ stand for the surface \Bf\  and the radius of the corresponding star, 
and $r_{em}$ is the distance from the star's surface.
Resolving (\ref{44}-\ref{441})  we find 
\ba &&
r _{em} = \frac{e }{ m_e ^{5/7} c^{11/7}} { B_{{WD}}^{3/7} R_{{WD} }^{9/7} \om ^{1/7} }
\nn &&
N_p = \frac{ B_{{WD}}^{2/7} R_{{WD} }^{6/7} L_s}{ c^{19/7} m_e^{8/7}  \om^{4/7}}
\nn &&
\gamma_{em}=  \frac{e }{ m_e ^{4/7} c^{13/7}} { B_{{WD}}^{1/7} R_{{WD} }^{3/7} \om ^{5/7} }
\label{em0}
\ea
Curiously, particles with very high energy radiate at higher frequencies further out. 

%Also, relations (\ref{em}) assume synchrotron emission, $\gamma \gg 1$ to translates to  $\om \gg  ( c^2 \sqrt{m_e}/e) /a^{1/2} =  2 \times 10^{11} $ rad s$^{-1}$.  {\color{red} I can't reproduce this number, why it depends on a?})

For optical  synchrotron emission with $ L_{o} \sim 10^{32}$ erg s$^{-1}$
\ba &&
r _{em} = 3.4\times10^{9} \;\om^{1/7}_{15} \mbox{ cm}
\nn &&
N_p = 3\times10^{34} \;  L_{o,32} \om^{-4/7}_{15}
\nn &&
\gamma_ {em}= 80 \; \om^{5/7}_{15}
\label{em}
\ea

%For particles propagation toward the WD and emitting in the optical $\om_o \approx 10^{15}$ rad s$^{-1}$, relations (\ref{em}) give 
%\ba &&
%r _{em,o}=5 \times 10^9 {\rm cm}
%\nn &&
%N_{p,o} =  10^{36}
%\nn &&
%\gamma_ {em,o}=  300
%\label{em}
%\ea
 Thus, relativistic  particles in the WD \ms\ emit optical emission at $ \approx 10 R_{WD}$. The amount of mass participating in the optical emission $ \sim m_p N_p\approx 5\times10^{10}$ g, is fairly small. The \Bf\ in the optical emission region evaluates to $B_{em}\approx 2500 \om^{-3/7}_{15} $~G. 
 %(We also note that the
% estimate of the \Bf\ (\ref{em}) is similar to the assumed  surface fields of the RD. Thus, one might expect optical emission to come both from the WD's and RD's \mss.)
 
  \cite{2018ApJ...853..106T}  reported observations of AR Sco in the X-ray range with luminosity $L_X \sim 4 \times 10^ {30} $ erg s$^{-1}$. Corresponding relations for X-rays give
\ba &&
r _{em,X} = 1.2\times10^{10} \;\om^{1/7}_{19} \mbox{ cm}
\nn &&
N_{p,X} = 2\times10^{32} \; \om^{-4/7}_{19}
\nn &&
\gamma_ {em,X}= 6\times10^{4} \; \om^{5/7}_{19}
\label{em2}
\ea
For $\om_X \approx 10^{19}$ rad s$^{-1}$.
 The spectrum of the accelerated particles corresponds to  $f \propto \gamma^{-p}$ with $p\approx 2$.  All very reasonable numbers.

% Equating the maximal \Lf\ (\ref{gammamax}) to the cooling limited one (\ref{em}), we conclude that the maximal expected frequency is
% \be
% \om_{max} = \frac{e^{1/2}}{ m_e c^{5/2}} \frac{B_{WD}^{3/2} R_{WD}^{9/2} \Omega_{WD} ^{3/2}}{a^2} = 4 
% \times 10^{20} {\rm rad s}^{-1}
% \ee
% (corresponding to $\sim 200$ keV). This is an  upper limit on the expected synchrotron  signal; it assumes the most efficient acceleration in reconnection.  

The inverse Compton scattering of electrons with \Lfs\ given in (\ref{em}) on the WD's photons with $\epsilon_{WD} \sim 1 $ eV would produce similar frequencies, $ \epsilon_{IC} \sim \gamma_{em}^2 \epsilon_{WD} \sim 10^{4} \, {\rm eV}$.  The electrons with the maximum \Lf\ (\ref{gammamax}) would produce IC emission in the TeV range. Unfortunately, a magnetic
energy density $U_B\approx B_{em}^2/8\pi \sim 10^7$ is much higher than the soft photon energy density $U_{ph}\sim L_o/4\pi c r_{em}^2\sim 30$, so for the leptonic model, the high energy emission can be estimated as $L_{HE}\sim L_o U_{ph}/U_B \sim 10^{26} $~erg/s. For a distance $\sim$100~pc we expect the observed flux to be $F_{HE}\sim10^{-16}$~erg/s~cm$^2$.  

In conclusion, optical and X-ray emission from the system originates due to synchrotron emission of particles accelerated in reconnection events. Synchrotron cooling determines the typical location and luminosity.

% {\color{red} We need estimations for radio from RD in cyclotron regime.}

\subsection{Radio emission}
\label{radio}

%\subsubsection{An analogue of Jupiter's decametric wavelengths emission in the RD's magnetosphere}
The radio emissions from AR Sco paint a murky picture of the underlying physical processes that cause them. 
\cite{2016Natur.537..374M}  found that AR Sco was a source of broadband, pulsed, $<$10\% circularly polarized radio emission at high brightness temperatures ($T_{B}\sim10^{12}-10^{14}$ K) for reasonable size estimates of the emitting region.
Karl G. Jansky Very Large Array (VLA) observations of the system detected non-thermal emissions that were 0 to -27\% circularly polarized on timescales of $\sim$10  min at 1.5 GHz, but only 0 to -8\% at 5 and 9 GHz \citep{2018A&A...611A..66S}.
The measured linear polarization fractions were small, totaling 0-3\%, and therefore much smaller than the degree of linear polarization measured at optical wavelengths.
This small linear polarization fraction cannot be explained by synchrotron radio emission from the WD magnetosphere alone, again suggesting that AR Sco is not a ``WD pulsar.''
 
We note that the observed properties of AR Sco bear many similarities to Jovian radio emission, as predicted by \cite{2004MNRAS.348..285W} in their work on the ECM-generated radio emissions anticipated to be found from planets that may orbit a WD.
In our own solar system, Jupiter is a bright radio source of decametric emission (DAM), see review by \cite{2017RvMPP...1....5M}. DAM is nearly 100\% circularly or elliptically polarized, and beamed in a hollow cone perpendicular to the source magnetic field, which indicates an ECM origin  \citep{1998JGR...10320159Z}.
DAM emission is also driven by the binary interaction of magnetospheres - in this case of Jupiter's \ms\ with Io \citep{1969ApJ...156...59G}.
In this example, the emission is produced by the electrons accelerated by Io's generated inductive \Ef.
Many of these observed properties qualitatively resemble the radio emission in AR Sco, which the present model also suggests is caused by the binary interaction of magnetospheres.   

An alternative explanation for AR Sco's radio emissions is gyrosynchrotron radiation.  The standard flare scenario on the Sun and other main sequence stars is that reconnection in their magnetospheres accelerate electrons that emit mildly relativistic gyrosynchrotron radiation at radio wavelengths, and hard X-rays via bremsstrahlung when they interact with the denser layers of the corona \citep{1998ARA&A..36..131B}.  This non-thermal emission mechanism extends to red dwarfs of spectral types as late as M9.  These ``ultracool dwarfs'' are sources of quiescent, non-bursty, radio emissions at 2-8 GHz with circular polarization fractions $<$35\% that have been the subject of extensive plasma physics modeling efforts \citep{2017MNRAS.465.1995M,2019MNRAS.483..614Z}.  Within the AR Sco system, low circular polarization fractions, coupled with larger potential source region sizes imply the operation of an incoherent emission process such as gyrosynchrotron radiation.
 
In addition to polarization fraction and brightness temperature measurements, another means to distinguish between these two emission mechanisms is to leverage the G\"{u}del-Benz relationship, which relates the thermalized X-ray luminosity generated by magnetic reconnection in stellar flares to the nonthermal, incoherent, gyrosynchrotron radio emission that results from particle acceleration (for a review, see \citet{2010ARA&A..48..241B}). This relationship is given by
%  {\color{red} I did not understand this and paragraph bellow, It looks like mistake in dimensions.}
%
\begin{equation} {L_{X} \over L_{R}}\sim 10^{15.5\pm 0.5}~[Hz],\end{equation} 
where $L_{X}$ is the X-ray luminosity and $L_{R}$ is the radio luminosity, usually computed as $\nu L_{\nu}$.  \citet{2016Natur.537..374M} measured an X-ray luminosity of $L_{X}\approx 4.9\times 10^{30}$ erg s$^{-1}$ using \emph{Swift}/X-ray Telescope (XRT), which corresponds to a G\"{u}del-Benz relationship peak radio luminosity $\nu L_{\nu}\sim 4.9\times 10^{15}$ erg s$^{-1}$. 

  In contrast to this low expected radio luminosity, \citet{2016Natur.537..374M} measured a peak radio flux density of $F_{\nu,peak}\sim 15$ mJy at 9 GHz with the Australian Telescope Compact Array (ATCA), which corresponds to a peak radio luminosity of $\nu L_{\nu}\sim 2.3\times 10^{27}$ erg s$^{-1}$.  This is far in excess of the radio emission generated by typical stellar flaring.  The combination of moderate circular polarization fractions coupled with a greater-than-expected radio luminosity given the X-ray activity within the system indicates that both gyrosynchrotron and ECM processes must be present, caused by the complex interactions of the two magnetospheres.

Unlike X-ray and optical electrons which are accelerated to relativistic energies within the WD magnetosphere, the radio electrons are not cooling  efficiently, hence estimates (\ref{44}) are not applicable.  The relatively large circular polarization fractions imply  mildly relativistic electrons at most. The frequency of $9$ GHz can then be used to estimate the \Bf\ on the RD:  $B_{RD} \sim 3\times 10^2 \gamma_{0.5}^2$ G, where we assume mildly relativistic electrons.
The number of radio emitting electrons then estimates to
\be
N_r \approx \frac{m_e^2 c^3 L_r}{e^4 B_{RD}^2} \approx 7\times 10^{35}  \gamma_{0.5}^2
\ee
We note that if ECM emission is present within the RD magnetosphere, there should exist a definite cutoff frequency, $\nu_{cf}$ beyond which radio emission is not detected that denotes the maximum magnetic field strength in the emitting region.  The presence of this spectral feature would enable the direct calculation of the magnetic field strength  and place constraints on the emitting plasma density \citep[\eg][]{2012ApJ...747L..22R}.

Additional insight can be gained through analysis of the radio flux variation within the AR Sco system on timescales on the order of an orbital period.  In Figure 4 of \citet{2018A&A...611A..66S}, the radio emissions from 1-10 GHz create a sinusoidal envelope with peak flux density $F_{\nu,peak}\sim 12$ mJy occurring near $\phi_{orb}\sim 0.5$, which corresponds to the WD being closest to the Earth.  Although the radio flux density decreases to $F_{\nu}\sim 5$ mJy at $\phi_{orb}\sim 0$, it does not disappear entirely.  This simple fact enables us to estimate that the RD contributes $\sim$40\% of the system's radio emission, while the remaining $\sim$60\% is generated by the reconnection emission model described in \S \ref{emit}.  Similarly, the orbital modulation of the circular polarization fraction presents clues as to the location where cyclotron emission may arise.  This fraction is maximal near $\phi_{orb}\sim -0.1$, when the observing geometry favors an unobstructed view of the RD hemisphere nearest the WD and the magnetospheric interaction region.  Thus, these considerations support our model of magnetospheric interactions causing the nonthermal acceleration of electrons, which in the RD magnetosphere, result in additional cyclotron emission superimposed on intrinsic RD stellar flaring.

% \cite{2018A&A...611A..66S} also measured a spectral index of $\alpha$=0.358 for non-simultaneous emissions from Ar Scorpii, which is similar to that found in optically-thick bremsstrahlung emission, as occurs, for instance, in YSOs (Dzib et al. 2013).  The spectral index is an interesting part: values of -1 indicate non-thermal emission, yet it is mildly positive.

% On the other hand, the $<$30\% circular polarization fraction appears to be inconsistent with the ECM scenario, which generally has $\sim$100\% polarization fractions.  Temporal averaging of bursty ECM emission and propagation effects may reduce the polarization fraction from ECM emission, particularly from regions near the limb.

% For reasonable size estimates for the RD magnetosphere, they computed $T_{B}\sim10^{12}-10^{14}$ K), or $T_{B}\sim10^{9}$ K for source region sizes supported by light travel time arguments.  8.5 GHz Australian Long Baseline Array (LBA) interferometry loosely constrained the radio emitting region size to be $\lesssim$4$R_{\odot}$, or three times the length of the system semi-major axis, with a non-thermal brightness temperature of $T_{B}\gtrsim 5\times 10^{9}$ K \citep{2017A&A...601L...7M}. 

% {\color{red} We did not get an answer how radio emission is formed.}
 
%\subsubsection{Radio emission: estimates}

\section{AE Aqr system and other polars}
\label{AEAqr}

AE Aqr \citep{1979ApJ...234..978P,1997MNRAS.286..436W,1994MNRAS.267..577D}  is  the most rapidly rotating white dwarf known ($P_{\rm rot} = 33.08$ s); it is  also the most strongly asynchronous object ($P_{orb} =9.88$ hr) in the DQ Herculis class. 
 AE~Aqr   is classified as a DQ Herculis-type cataclysmic variable, comprising a magnetized white dwarf primary and a K5 dwarf secondary.  AE Aqr is characterized by coherent pulsations and quasi-periodic oscillations (QPOs) in the optical, UV, and soft X-ray wavelength bands.  Although early work suggested that it is a source of 0.35-2.4 TeV $\gamma$-rays, later results from MAGIC and the \emph{FERMI} Large Area Telescope (LAT) failed to confirm these purported detections \citep{1992ApJ...401..325M,2014A&A...568A.109A,2016ApJ...832...35L}. In addition, AE Aqr displays violent flaring activity at optical, soft X-ray, and radio wavelengths.
 
 We note in particular how the radio emission from AE Aqr differs from that found from AR Sco.  Non-simultaneous, VLA observations of AE Aqr at 1.4, 4.9, 15, and 22.5 GHz revealed radio emission that varied on timescales of $\sim$5 min, with greater variability detected at higher frequencies \citep{1987ApJ...323L.131B,1988ApJ...324..431B}.  No circular polarization was detected to within instrumental uncertainty ($\lesssim$15\%) and the spectral index was found to vary from $\alpha\sim-1$ to 1.5.  These results led \citet{1987ApJ...323L.131B} to suggest that synchrotron radiation from a mildly relativistic population of electrons ($\gamma\sim3$) caused the radio emission, with the WD acting as an injector of electrons that are confined within the magnetic bottle of the secondary's strong magnetic field.  Alternatively, \citet{1988ApJ...324..431B} argued that the radio emission represented the superposition of almost continually occurring synchrotron flares.

Let us apply the model to the AE Aqr.
 From (\ref{Rc}) the required  accretion rate during the spin-up stage is
 \be
 \dot{M} = 4\pi \frac{B_{WD}^2 R_{WD}^6 \Omega_{WD}^{7/3}}{ ( G M_{WD})^{5/3}}= 
 \left\{
 \begin{array}{cc}
 10^{-7}  B_{WD, 6}^2 & \mbox{for AR Sco}\\
 10^{-6}  B_{WD, 6} ^2 & \mbox{for AE Aqr}
 \end{array}
 \right.
 \label{MMdot}
 \ee
 Thus, smaller $\dot{M} $ is required for AE Aqr during the high stage; equivalently, its \Bf\ can be somewhat higher. 
Our conclusion about the properties of AE Aqr are, generally, in agreement with those reached by \cite{2019MNRAS.487.1754B}.

Thus, the model places AR Sco  (and AE Aqr) within a short-lived phase of IPs,  with a very IP-like magnetic field strength.  The AR Sco stage is short, $\sim 10^6-10^7$ yrs. Since, typically, the IP phase lasts $\sim 10^9$  years,  several transitions to such a state can occur during the system's lifetime.

 What distinguishes AR Sco  and AE Aqr from other intermediate   polars?  IPs typically have  \Bf\ $\sim 10^7$ G and are accreting. We suggest: (i)
AR Sco  and AE Aqr have  {\it smaller} \Bfs: the equilibrium spin is inversely proportional to the surface \Bf\  (for a given $\dot{M}$) -  from (\ref{Rc}) $\Omega \propto \dot{M}^{3/7} B_{WD}^{-6/7}$. Thus, small WD surface fields allow for faster equilibrium spin during the spin-up stage \citep[low \Bf\  in AE Aqr was also proposed by][]{2004PASP..116..115W}; (ii) 
  currently AR Sco  and AE Aqr  are in a propeller regime due to their low accretion rates.

The present model also can be related to the ``hibernating intermediate polar'' model of \cite{2002AIPC..637....3W}, which proposes that   the mass lost by the WD during a nova will cause the secondary to detach from its Roche lobe. The mass-transfer rate  then drops to extremely low  levels for very long periods of time. But before the system enters hibernation, there is a brief  interval of enhanced mass transfer, caused by the irradiation of the secondary.  Thus, various  states of the system would involve:  (i)  high $\dot{M}$, accretion, spin-up; (ii) nova explosion, ejection of material; (iii) on Kelvin time scales the companion relaxes to a new detached state,  and becomes a hibernating intermediate polar with very small $\dot{M}$.

\section{Discussion}

We develop   a model of     the  highly asynchronous intermediate polars AR Sco and AE Aqr.  The \Bfs\  of the WDs  are   relatively weak, $\sim 10^7$ G.  They are currently in a transient propeller stage. The weak \Bfs\ allowed a system to be in the accretor state during previous high mass transfer stage. As the WDs are spinning down quickly, each will eventually be in a double synchronous state like AM Herculis \citep{1979ApJ...230..176J}. 
The propeller stage in AR Sco does not even involve formation of the disk, but direct magnetospheric interaction with the companion.  
The  fast spin-down of the WDs is  determined by  loading of the RD's material and ensuing expulsion from the WDs' \mss\ in a transient  propeller regime. 
If the mass accretion rate remains small, as it is now, each system will become synchronous and eventually will start accreting \citep[this regime was  studied numerically by][]{2012PhyU...55..115Z,2019INASR...3..194I,2019ARep...63..751Z}.  But if $\dot{M}$ increases, they will enter the earlier accretor regime.

In both systems, the mass loading of the WD's \ms\ by the partially ionized  RD's stream is strongly affected by the ionizing  radiation from the WD. It leads to efficient loading of the WD's \ms\ needed to explain the high spin-down rate. The ionization conditions, we hypothesize, are what make the AR Sco  and AE Aqr different:  in our model the ionization of the RD's flow by the WD is important, this difference in temperatures might affect the flow dynamics, as discussed at the end of  Section 3.2.

%\textbf{we should discuss it more.}

We envision that most of the observed properties are determined by the direct interaction of the stars' \mss. This requires that the \Alfven points in the corresponding winds are further way from the stars than the $L_1$ point.  This is easily achieved for the RD, since it is almost Roche lobe filling; 
in the case of WD it is required the the \Alfven velocity in the \ms\ is larger that $v_A \geq (a \Omega_{WD}) \approx 0.1 c$.
The  magnetospheric/wind interaction  of two stars  is not responsible for the WD's spin-down:   it  leads to the generation of the observed nonthermal emission by particles accelerated in reconnection events.
% For radio electrons cooling is not important, even closed to the WD's surface. 

Finally, we point out that  conventional models of spin and orbital evolution may have to be corrected  in the case of   AR Sco and AE Aqr. 
Interaction of the WD's and RD's \mss\ also lead to a torque on the RD \citep{1967AcA....17..287P,1981A&A...100L...7V}. Changing the  spin of the RD, combined with the spin-orbital tidal synchronization, and the corresponding loss of the orbital angular  momentum and the size of the  RD's Roche lobe, will lead to changes in the mass accretion rate. 
Using (\ref{Lsd11}), we can estimate  the mutual torque as
\be 
\dot{J}  =  -  \frac{1}{4} \frac{\mu_{WD} \mu_{RD}} {a^3}= 10^{33}  {\rm erg} 
\label{J}
\ee
for the parameters of AR Sco. This is the torque exerted on the RD due to magnetospheric interaction with the WD. 
This comes close to the general relativistic  torque, which for the parameters of  AR Sco,  evaluates to $10^{34}$ erg. (In fact, Eq. (\ref{J}) underestimates the torque, since it is applied at $r_{int}  < a$.) We leave consideration of these effects to a subsequent paper.

   %%%%%%%%%%%%%%%%%%%%%%%%%%%%%%%%%%%%%%%%%%%%%%%%%%%%%%%%%%%%%%%%%
\section*{Acknowledgments}
This work had been supported by DoE grant DE-SC0016369,
NASA grant 80NSSC17K0757 and  NSF grants 10001562 and 10001521. 
ML would like to thank organizers and participants of the conference ``Compact White Dwarf Binaries'' for enlightening discussions. We also thank  David Buckley, Paul Callanan, Nazar Ikhsanov and  Thomas Marsh for the most valuable comments.  MR acknowledges that this research was supported in part through computational resources provided by Information Technology at Purdue, West Lafayette, IN.
%%%%%%%%%%%%%%%%%%%%%%%%%%%%%%%%%%%%%%%%%%%%%%%%%%%%%%%%%%%%%%%%%

%\item MR: I've merged refs from 4 differnt .bbl files into the bibliography, but still noted some refs were unaccounted for.  The .bbl file that corresponds to this tex file needs to also be examined to see if there are any unused refs.

\bibliographystyle{apj}
  \bibliography{/Users/maxim/Home/Research/BibTex,arsco}

\begin{thebibliography}{59}
\expandafter\ifx\csname natexlab\endcsname\relax\def\natexlab#1{#1}\fi

\bibitem[{{Abdo} {et~al.}(2013){Abdo}, {Ajello}, {Allafort}, {Baldini},
  {Ballet}, {Barbiellini}, {Baring}, {Bastieri}, {Belfiore}, {Bellazzini}, \&
  et~al.}]{2013ApJS..208...17A}
{Abdo}, A.~A., {et~al.} 2013, \apjs, 208, 17

\bibitem[{{Aleksi{\'c}} {et~al.}(2014){Aleksi{\'c}}, {Ansoldi}, {Antonelli},
  {Antoranz}, {Babic}, {Bangale}, {Barrio}, {Becerra Gonz{\'a}lez}, {Bednarek},
  {Bernardini}, {Biasuzzi}, {Biland}, {Blanch}, {Bonnefoy}, {Bonnoli},
  {Borracci}, {Bretz}, {Carmona}, {Carosi}, {Colin}, {Colombo}, {Contreras},
  {Cortina}, {Covino}, {Da Vela}, {Dazzi}, {De Angelis}, {De Caneva}, {De
  Lotto}, {de O{\~n}a Wilhelmi}, {Delgado Mendez}, {Doert}, {Dominis Prester},
  {Dorner}, {Doro}, {Einecke}, {Eisenacher}, {Elsaesser}, {Fonseca}, {Font},
  {Frantzen}, {Fruck}, {Galindo}, {Garc{\'\i}a L{\'o}pez}, {Garczarczyk},
  {Garrido Terrats}, {Gaug}, {Godinovi{\'c}}, {Gonz{\'a}lez Mu{\~n}oz},
  {Gozzini}, {Hadasch}, {Hanabata}, {Hayashida}, {Herrera}, {Hildebrand },
  {Hose}, {Hrupec}, {Idec}, {Kadenius}, {Kellermann}, {Kodani}, {Konno},
  {Krause}, {Kubo}, {Kushida}, {La Barbera}, {Lelas}, {Lewandowska},
  {Lindfors}, {Lombardi}, {L{\'o}pez}, {L{\'o}pez-Coto}, {L{\'o}pez-Oramas},
  {Lorenz}, {Lozano}, {Makariev}, {Mallot}, {Maneva}, {Mankuzhiyil},
  {Mannheim}, {Maraschi}, {Marcote}, {Mariotti}, {Mart{\'\i}nez}, {Mazin},
  {Menzel}, {Mirand a}, {Mirzoyan}, {Moralejo}, {Munar-Adrover}, {Nakajima},
  {Niedzwiecki}, {Nilsson}, {Nishijima}, {Noda}, {Nowak}, {Orito},
  {Overkemping}, {Paiano}, {Palatiello}, {Paneque}, {Paoletti}, {Paredes},
  {Paredes-Fortuny}, {Persic}, {Prada Moroni}, {Prandini}, {Preziuso},
  {Puljak}, {Reinthal}, {Rhode}, {Rib{\'o}}, {Rico}, {Rodriguez Garcia},
  {R{\"u}gamer}, {Saggion}, {Saito}, {Saito}, {Satalecka}, {Scalzotto},
  {Scapin}, {Schultz}, {Schweizer}, {Sillanp{\"a}{\"a}}, {Sitarek}, {Snidaric},
  {Sobczynska}, {Spanier}, {Stamatescu}, {Stamerra}, {Steinbring}, {Storz},
  {Strzys}, {Takalo}, {Takami}, {Tavecchio}, {Temnikov}, {Terzi{\'c}},
  {Tescaro}, {Teshima}, {Thaele}, {Tibolla}, {Torres}, {Toyama}, {Treves},
  {Uellenbeck}, {Vogler}, {Wagner}, \& {Zanin}}]{2014A&A...568A.109A}
{Aleksi{\'c}}, J., {et~al.} 2014, \aap, 568, A109

\bibitem[{{Barlow} {et~al.}(2006){Barlow}, {Knigge}, {Bird}, {J Dean}, {Clark},
  {Hill}, {Molina}, \& {Sguera}}]{2006MNRAS.372..224B}
{Barlow}, E.~J., {Knigge}, C., {Bird}, A.~J., {J Dean}, A., {Clark}, D.~J.,
  {Hill}, A.~B., {Molina}, M., \& {Sguera}, V. 2006, \mnras, 372, 224

\bibitem[{{Barrett} {et~al.}(2017){Barrett}, {Dieck}, {Beasley}, {Singh}, \&
  {Mason}}]{2017AJ....154..252B}
{Barrett}, P.~E., {Dieck}, C., {Beasley}, A.~J., {Singh}, K.~P., \& {Mason},
  P.~A. 2017, \aj, 154, 252

\bibitem[{{Bastian} {et~al.}(1998){Bastian}, {Benz}, \&
  {Gary}}]{1998ARA&A..36..131B}
{Bastian}, T.~S., {Benz}, A.~O., \& {Gary}, D.~E. 1998, \araa, 36, 131

\bibitem[{{Bastian} {et~al.}(1988){Bastian}, {Dulk}, \&
  {Chanmugam}}]{1988ApJ...324..431B}
{Bastian}, T.~S., {Dulk}, G.~A., \& {Chanmugam}, G. 1988, \apj, 324, 431

\bibitem[{{Benz} \& {G{\"u}del}(2010)}]{2010ARA&A..48..241B}
{Benz}, A.~O., \& {G{\"u}del}, M. 2010, \araa, 48, 241

\bibitem[{{Beuermann}(1999)}]{1999hxra.conf..410B}
{Beuermann}, K. 1999, in Highlights in X-ray Astronomy, ed. B.~{Aschenbach} \&
  M.~J. {Freyberg}, Vol. 272, 410

\bibitem[{{Blinova} {et~al.}(2019){Blinova}, {Romanova}, {Ustyugova},
  {Koldoba}, \& {Lovelace}}]{2019MNRAS.487.1754B}
{Blinova}, A.~A., {Romanova}, M.~M., {Ustyugova}, G.~V., {Koldoba}, A.~V., \&
  {Lovelace}, R.~V.~E. 2019, \mnras, 487, 1754

\bibitem[{{Bookbinder} \& {Lamb}(1987)}]{1987ApJ...323L.131B}
{Bookbinder}, J.~A., \& {Lamb}, D.~Q. 1987, \apjl, 323, L131

\bibitem[{{Buckley} {et~al.}(2017){Buckley}, {Meintjes}, {Potter}, {Marsh}, \&
  {G{\"a}nsicke}}]{2017NatAs...1E..29B}
{Buckley}, D.~A.~H., {Meintjes}, P.~J., {Potter}, S.~B., {Marsh}, T.~R., \&
  {G{\"a}nsicke}, B.~T. 2017, Nature Astronomy, 1, 0029

\bibitem[{{de Jager} {et~al.}(1994){de Jager}, {Meintjes}, {O'Donoghue}, \&
  {Robinson}}]{1994MNRAS.267..577D}
{de Jager}, O.~C., {Meintjes}, P.~J., {O'Donoghue}, D., \& {Robinson}, E.~L.
  1994, \mnras, 267, 577

\bibitem[{{du Plessis} {et~al.}(2019){du Plessis}, {Wadiasingh}, {Venter}, \&
  {Harding}}]{2019arXiv191007401D}
{du Plessis}, L., {Wadiasingh}, Z., {Venter}, C., \& {Harding}, A.~K. 2019,
  arXiv e-prints, arXiv:1910.07401

\bibitem[{{Eggleton}(1983)}]{1983ApJ...268..368E}
{Eggleton}, P.~P. 1983, \apj, 268, 368

\bibitem[{{Fawley} {et~al.}(1977){Fawley}, {Arons}, \&
  {Scharlemann}}]{1977ApJ...217..227F}
{Fawley}, W.~M., {Arons}, J., \& {Scharlemann}, E.~T. 1977, \apj, 217, 227

\bibitem[{{Ferrario} {et~al.}(2015){Ferrario}, {de Martino}, \&
  {G{\"a}nsicke}}]{2015SSRv..191..111F}
{Ferrario}, L., {de Martino}, D., \& {G{\"a}nsicke}, B.~T. 2015, \ssr, 191, 111

\bibitem[{{Garnavich} {et~al.}(2019){Garnavich}, {Littlefield}, {Kafka},
  {Kennedy}, {Callanan}, {Balsara}, \& {Lyutikov}}]{2019ApJ...872...67G}
{Garnavich}, P., {Littlefield}, C., {Kafka}, S., {Kennedy}, M., {Callanan}, P.,
  {Balsara}, D.~S., \& {Lyutikov}, M. 2019, \apj, 872, 67

\bibitem[{{Goldreich} \& {Julian}(1969)}]{GoldreichJulian}
{Goldreich}, P., \& {Julian}, W.~H. 1969, \apj, 157, 869

\bibitem[{{Goldreich} \& {Lynden-Bell}(1969)}]{1969ApJ...156...59G}
{Goldreich}, P., \& {Lynden-Bell}, D. 1969, \apj, 156, 59

\bibitem[{{Ikhsanov}(1998)}]{1998A&A...338..521I}
{Ikhsanov}, N.~R. 1998, \aap, 338, 521

\bibitem[{{Ikhsanov} \& {Beskrovnaya}(2008)}]{2008arXiv0809.1169I}
{Ikhsanov}, N.~R., \& {Beskrovnaya}, N.~G. 2008, arXiv e-prints,
  arXiv:0809.1169

\bibitem[{{Ikhsanov} \& {Beskrovnaya}(2012)}]{2012ARep...56..595I}
---. 2012, Astronomy Reports, 56, 595

\bibitem[{{Ikhsanov} \& {Biermann}(2006)}]{2006A&A...445..305I}
{Ikhsanov}, N.~R., \& {Biermann}, P.~L. 2006, \aap, 445, 305

\bibitem[{{Isakova} {et~al.}(2019){Isakova}, {Zhilkin}, \&
  {Bisikalo}}]{2019INASR...3..194I}
{Isakova}, P.~B., {Zhilkin}, A.~G., \& {Bisikalo}, D.~V. 2019, INASAN Science
  Reports, 3, 194

\bibitem[{{Joss} {et~al.}(1979){Joss}, {Katz}, \&
  {Rappaport}}]{1979ApJ...230..176J}
{Joss}, P.~C., {Katz}, J.~I., \& {Rappaport}, S. 1979, \apj, 230, 176

\bibitem[{{Kaplan} {et~al.}(2019){Kaplan}, {Meintjes}, {Singh}, {van Heerden},
  {Ramamonjisoa}, \& {van der Westhuizen}}]{2019arXiv190800283K}
{Kaplan}, Q., {Meintjes}, P.~J., {Singh}, K.~K., {van Heerden}, H.~J.,
  {Ramamonjisoa}, F.~A., \& {van der Westhuizen}, I.~P. 2019, arXiv e-prints,
  arXiv:1908.00283

\bibitem[{{Katz}(2017)}]{2017ApJ...835..150K}
{Katz}, J.~I. 2017, \apj, 835, 150

\bibitem[{{Knigge} {et~al.}(2011){Knigge}, {Baraffe}, \&
  {Patterson}}]{2011ApJS..194...28K}
{Knigge}, C., {Baraffe}, I., \& {Patterson}, J. 2011, \apjs, 194, 28

\bibitem[{{Koester}(2010)}]{2010MmSAI..81..921K}
{Koester}, D. 2010, \memsai, 81, 921

\bibitem[{{Li} {et~al.}(2016){Li}, {Torres}, {Rea}, {de O{\~n}a Wilhelmi},
  {Papitto}, {Hou}, \& {Mauche}}]{2016ApJ...832...35L}
{Li}, J., {Torres}, D.~F., {Rea}, N., {de O{\~n}a Wilhelmi}, E., {Papitto}, A.,
  {Hou}, X., \& {Mauche}, C.~W. 2016, \apj, 832, 35

\bibitem[{{Lyutikov} {et~al.}(2017{\natexlab{a}}){Lyutikov}, {Sironi},
  {Komissarov}, \& {Porth}}]{2017JPlPh..83f6301L}
{Lyutikov}, M., {Sironi}, L., {Komissarov}, S.~S., \& {Porth}, O.
  2017{\natexlab{a}}, Journal of Plasma Physics, 83, 635830601

\bibitem[{{Lyutikov} {et~al.}(2017{\natexlab{b}}){Lyutikov}, {Sironi},
  {Komissarov}, \& {Porth}}]{2017JPlPh..83f6302L}
---. 2017{\natexlab{b}}, Journal of Plasma Physics, 83, 635830602

\bibitem[{{Lyutikov} \& {Thompson}(2005)}]{2005ApJ...634.1223L}
{Lyutikov}, M., \& {Thompson}, C. 2005, \apj, 634, 1223

\bibitem[{{Marsh} {et~al.}(2016){Marsh}, {G{\"a}nsicke}, {H{\"u}mmerich},
  {Hambsch}, {Bernhard}, {Lloyd}, {Breedt}, {Stanway}, {Steeghs}, {Parsons},
  {Toloza}, {Schreiber}, {Jonker}, {van Roestel}, {Kupfer}, {Pala}, {Dhillon},
  {Hardy}, {Littlefair}, {Aungwerojwit}, {Arjyotha}, {Koester}, {Bochinski},
  {Haswell}, {Frank}, \& {Wheatley}}]{2016Natur.537..374M}
{Marsh}, T.~R., {et~al.} 2016, \nat, 537, 374

\bibitem[{{Meintjes} {et~al.}(1992){Meintjes}, {Raubenheimer}, {de Jager},
  {Brink}, {Nel}, {North}, {van Urk}, \& {Visser}}]{1992ApJ...401..325M}
{Meintjes}, P.~J., {Raubenheimer}, B.~C., {de Jager}, O.~C., {Brink}, C.,
  {Nel}, H.~I., {North}, A.~R., {van Urk}, G., \& {Visser}, B. 1992, \apj, 401,
  325

\bibitem[{{Melrose}(2017)}]{2017RvMPP...1....5M}
{Melrose}, D.~B. 2017, Reviews of Modern Plasma Physics, 1, 5

\bibitem[{{Metodieva} {et~al.}(2017){Metodieva}, {Kuznetsov}, {Antonova},
  {Doyle}, {Ramsay}, \& {Wu}}]{2017MNRAS.465.1995M}
{Metodieva}, Y.~T., {Kuznetsov}, A.~A., {Antonova}, A.~E., {Doyle}, J.~G.,
  {Ramsay}, G., \& {Wu}, K. 2017, \mnras, 465, 1995

\bibitem[{{Paczy{\'n}ski}(1967)}]{1967AcA....17..287P}
{Paczy{\'n}ski}, B. 1967, \actaa, 17, 287

\bibitem[{{Patterson}(1979)}]{1979ApJ...234..978P}
{Patterson}, J. 1979, \apj, 234, 978

\bibitem[{{Radhakrishnan} \& {Cooke}(1969)}]{1969ApL.....3..225R}
{Radhakrishnan}, V., \& {Cooke}, D.~J. 1969, \aplett, 3, 225

\bibitem[{{Rees} \& {Gunn}(1974)}]{reesgunn}
{Rees}, M.~J., \& {Gunn}, J.~E. 1974, MNRAS, 167, 1

\bibitem[{{Route} \& {Wolszczan}(2012)}]{2012ApJ...747L..22R}
{Route}, M., \& {Wolszczan}, A. 2012, \apjl, 747, L22

\bibitem[{{Stanway} {et~al.}(2018){Stanway}, {Marsh}, {Chote}, {G{\"a}nsicke},
  {Steeghs}, \& {Wheatley}}]{2018A&A...611A..66S}
{Stanway}, E.~R., {Marsh}, T.~R., {Chote}, P., {G{\"a}nsicke}, B.~T.,
  {Steeghs}, D., \& {Wheatley}, P.~J. 2018, \aap, 611, A66

\bibitem[{{Sturrock}(1971)}]{Sturrock71}
{Sturrock}, P.~A. 1971, \apj, 164, 529

\bibitem[{{Takata} {et~al.}(2018){Takata}, {Hu}, {Lin}, {Tam}, {Pal}, {Hui},
  {Kong}, \& {Cheng}}]{2018ApJ...853..106T}
{Takata}, J., {Hu}, C.~P., {Lin}, L.~C.~C., {Tam}, P.~H.~T., {Pal}, P.~S.,
  {Hui}, C.~Y., {Kong}, A.~K.~H., \& {Cheng}, K.~S. 2018, \apj, 853, 106

\bibitem[{{Toloza} {et~al.}(2019){Toloza}, {Breedt}, {De Martino}, {Drake},
  {Gansicke}, {Green}, {Ederoclite}, {Jennifer}, {Juna}, {Knigge}, {Kupfer},
  {Long}, {Marsh}, {Pala}, {Parsons}, {Prince}, {Raddi}, {Rebassa-Manserga},
  {Rodriguez-Gil}, {Scaringi}, {Schmidtobreick}, {Schreiber}, {Schwope},
  {Shen}, {Steeghs}, {Szkody}, {Tappert}, {Toonen}, {Townsley}, \&
  {Zorotovic}}]{2019BAAS...51c.168T}
{Toloza}, O., {et~al.} 2019, \baas, 51, 168

\bibitem[{{Usov}(1988)}]{1988SvAL...14..258U}
{Usov}, V.~V. 1988, Soviet Astronomy Letters, 14, 258

\bibitem[{{Verbunt} \& {Zwaan}(1981)}]{1981A&A...100L...7V}
{Verbunt}, F., \& {Zwaan}, C. 1981, \aap, 100, L7

\bibitem[{{Warner}(1995)}]{1995ASPC...85....3W}
{Warner}, B. 1995, Astronomical Society of the Pacific Conference Series,
  Vol.~85, {The Discovery Of Magnetic Cataclysmic Variable Stars}, ed. D.~A.~H.
  {Buckley} \& B.~{Warner}, 3

\bibitem[{{Warner}(2002)}]{2002AIPC..637....3W}
{Warner}, B. 2002, in American Institute of Physics Conference Series, Vol.
  637, Classical Nova Explosions, ed. M.~{Hernanz} \& J.~{Jos{\'e}}, 3--15

\bibitem[{{Warner}(2003)}]{2003cvs..book.....W}
---. 2003, {Cataclysmic Variable Stars}

\bibitem[{{Warner}(2004)}]{2004PASP..116..115W}
---. 2004, \pasp, 116, 115

\bibitem[{{Wickramasinghe} \& {Ferrario}(2000)}]{2000PASP..112..873W}
{Wickramasinghe}, D.~T., \& {Ferrario}, L. 2000, \pasp, 112, 873

\bibitem[{{Willes} \& {Wu}(2004)}]{2004MNRAS.348..285W}
{Willes}, A.~J., \& {Wu}, K. 2004, \mnras, 348, 285

\bibitem[{{Wynn} {et~al.}(1997){Wynn}, {King}, \&
  {Horne}}]{1997MNRAS.286..436W}
{Wynn}, G.~A., {King}, A.~R., \& {Horne}, K. 1997, \mnras, 286, 436

\bibitem[{{Zarka}(1998)}]{1998JGR...10320159Z}
{Zarka}, P. 1998, \jgr, 103, 20159

\bibitem[{{Zhilkin} {et~al.}(2012){Zhilkin}, {Bisikalo}, \&
  {Boyarchuk}}]{2012PhyU...55..115Z}
{Zhilkin}, A.~G., {Bisikalo}, D.~V., \& {Boyarchuk}, A. r.~A. 2012, Physics
  Uspekhi, 55, 115

\bibitem[{{Zhilkin} {et~al.}(2019){Zhilkin}, {Sobolev}, {Bisikalo}, \&
  {Gabdeev}}]{2019ARep...63..751Z}
{Zhilkin}, A.~G., {Sobolev}, A.~V., {Bisikalo}, D.~V., \& {Gabdeev}, M.~M.
  2019, Astronomy Reports, 63, 751

\bibitem[{{Zic} {et~al.}(2019){Zic}, {Lynch}, {Murphy}, {Kaplan}, \&
  {Chandra}}]{2019MNRAS.483..614Z}
{Zic}, A., {Lynch}, C., {Murphy}, T., {Kaplan}, D.~L., \& {Chandra}, P. 2019,
  \mnras, 483, 614

\end{thebibliography}

\end{document}